\documentstyle[11pt]{article}
\newcommand{\be}{\begin{eqnarray}}
\newcommand{\ee}{\end{eqnarray}}
\newcommand{\ra}{\rightarrow}

\begin{document}
\hskip 12.0cm
 KAIST-CHEP-96/07

\begin{center}
{\Large \bf Form Factors for Exclusive Semileptonic $B$--Decays }
\vskip 1.0cm
{ C. S. ~Kim\footnote{kim@cskim.yonsei.ac.kr,~~~ cskim@theory.kek.jp}
\\
\it Department of Physics, Yonsei University, Seoul 120--749, KOREA\\
{\rm and}\\
Theory Division, KEK, Tsukuba, Ibaraki 305, JAPAN}
\vskip 0.5cm
{ Jae Kwan ~Kim,~ Yeong Gyun ~Kim\footnote{ygkim@chep6.kaist.ac.kr}~ and~
Kang Young ~Lee\footnote{kylee@chep5.kaist.ac.kr,~~~kylee@prtcl1.yonsei.ac.kr}
\footnote{ Present address : 
Department of Physics, Yonsei University, Seoul 120--749, KOREA}
\\
\it Department of Physics, KAIST, Taejon 305--701, KOREA}
\end{center}

\begin{abstract}
We investigate the form factors for exclusive semileptonic 
decays of $B$-meson to $D,~D^*$, based on the parton picture 
and helped by the results of the HQET.  
We obtain the numerical results for the slope of the 
Isgur-Wise function, which is consistent
with the experimental results, and 
we extracte the dependences of hadronic form factors 
on $q^2$ by varying input heavy quark fragmentation function
without the nearest pole dominance ans\"atze. \\

\center{({\bf To be published by Int. J. Mod. Phys. A (1996).})}
\end{abstract}


\section{Introduction}

$B$--meson decay processes have been studied in detail
as providing many interesting informations on the interplay
of electroweak and strong interactions and as a source extracting 
the parameters of weak interactions, such as $|V_{cb}|$ and $|V_{ub}|$. 
As more data will be accumulated from the asymmetric $B$--factories 
in near future,
the theoretical and experimental studies on $B$--meson decays
would also give better understandings on the Standard Model and its 
possible extensions.

In exclusive weak decay processes of hadrons, the effects of strong
interaction are encoded in hadronic form factors.
These decay form factors are Lorentz invariant
functions which depend on the momentum transfer $q^2$,
and their behaviors with varying $q^2$ are dominated by non-perturbative
effects of QCD.

Over the past few years, a great progress has been achieved 
in our understanding of the exclusive semileptonic decays 
of heavy flavors to heavy flavors \cite{neubert}.
In the limit where the mass of the heavy quark is taken to infinity, 
its strong interactions become independent of its mass and spin, 
and depend only on its velocity. 
This provides a new $SU(2N_f)$ spin--flavor symmetry, 
which is not manifest in the theory of QCD. However, 
this new symmetry has been made explicit 
in a framework of the heavy quark effective theory (HQET) \cite{hqet}. 
In practice, the HQET and this new symmetry relate 
all the hadronic matrix elements of 
$B \ra D$ and $B \ra D^*$ semileptonic decays, 
and all the form factors can be reduced to a single universal function, 
the so-called Isgur-Wise function \cite{hqet,isgurfun},
which represents the common non-perturbative dynamics of 
weak decays of heavy mesons. 
However, the HQET cannot predict the values of the Isgur-Wise function 
over the whole $q^2$ range, though the normalization of the Isgur-Wise
function is precisely known in the zero recoil limit.
Hence the extrapolation of $q^2$ dependences of the Isgur-Wise function 
and of all form factors is still model dependent and the source of 
uncertainties in any theoretical studies. 
Therefore, it is strongly recommended to determine
hadronic form factors of $B$--meson decay more reliably, when we think
of their importance in theoretical and experimental analyses.

In this paper we develop a new approach for exclusive 
semileptonic $B$ decays to $D,~D^*$, and predict the $q^2$ dependences
of all form factors, inspired by the parton model. 
Previously the parton model approach has been 
established to describe inclusive semileptonic $B$ decays \cite{pas,jin},
and found to give excellent agreements with experiments for electron
energy spectrum at all energies.
The parton model approach for the inclusive decays 
pictures the mesonic decay as the decay of the partons
in analogy to deep inelastic scattering process.
The probability of finding a $b$-quark in a $B$ meson 
carrying a fraction $x$ of the meson momentum 
in the infinite momentum frame is given by the distribution function $f(x)$.

While many attempts describing exclusive $B$ decays often take
the pole-dominance ans$\ddot{\mbox {a}}$tze
as behaviors of form factors with varying $q^2$,
in our approach they are derived by the kinematical relations
between initial $b$ quark and final $c$ quark.
According to the Wirbel {\it et al.} model \cite{wsb}, 
which is one of the most popular model 
to describe exclusive decays of $B$ mesons, 
the hadronic form factors are related to
the meson wavefunctions' overlap-integral in the infinite momentum
frame, but in our model they are determined by integral 
of the distribution functions of quarks inside the mesons.

In Section II, we develop the parton model approach 
for exclusive semileptonic decays of $B$ meson. 
All the theoretical details for $B \ra D l \nu$ and $B \ra D^* l \nu$
are given in Section III. 
Section IV contains discussions and conclusions of this paper.

\section{Parton Model Approach for Exclusive Decays of B Meson}

We now develop the parton model approach for 
exclusive semileptonic decays of $B$ meson by 
extending the previously established inclusive parton model, 
and by combining with the results of the HQET.
Theoretical formulation of this approach is, in a sence,  related to
Drell-Yan process, while the parton model of inclusive $B$ decays is
motivated by deep inelastic scattering process.
And the bound state effects of exclusive $B$ decays are encoded into 
the hadronic distribution functions of partons inside an initial $B$ meson
and of partons of a final state resonance hadron.
In Fig. 1, we show the schematic diagrams of
Drell-Yan process and the related exlusive semileptonic decay of $B$ meson.
We assume that the Lorentz invariant hadronic decay width can be obtained
using the structure functions, 
\be
E_B \cdot d\Gamma(B \ra D(D^*)e\nu) = \int dx \int dy~
                      f_B(x)~E_b \cdot d\Gamma(b \ra ce\nu)~f_D(y)
\ee
following the Drell-Yan case.
The first integral represents the effects of motion of $b$ quark
within $B$ meson and the second integral those of $c$ quark within $D$ meson.
The variables $x$ and $y$ are defined as fractions of momenta of partons 
to momenta of mesons,
\be
p_b = x p_B~,~~~~~~~~~ 
p_c = y p_D~,
\ee
in the infinite momentum frame, $|{\bf p}| \gg m$.
In this frame, the eq. (2) is valid for the four momenta of 
massive partons.
The functions $f_B(x)$ and  $f_D(y)$ are the distribution function 
of $b$ quark inside $B$ meson, and the fragmentation function of $c$ quark 
to $D$ meson, respectively, which are also defined in that frame.
Since the momentum fractions and the distribution functions are 
all defined in the infinite momentum frame, we have to calculate the
Lorentz invariant quantity, ~$E \cdot d\Gamma$ as defined in Eq. (1), 
to use at any other frame safely.

The distribution function can be identified with the fragmentation function
for a fast moving $b$-quark to hadronize into a $B$ meson in the infinite
momentum frame.
In general the distribution and fragmentation functions 
of a heavy quark ($Q=t,b,c$) in a heavy meson ($Q q$),
which are closely related by a time reversal transformation,
are of very similar functional forms, and peak both at large value of $x$.
Brodsky {\it et al.} \cite{brod} have calculated 
the distribution function of a heavy quark,
which has the same form as Peterson's fragmentation function \cite{peterson}.
Therefore, here we follow the previous 
work of Paschos {\it et al.} \cite{pas}
to use the Peterson's fragmentation function for both
distributions,  $f_B(x)$ and  $f_D(y)$. It has the functional form:
\be
f_Q(z) = N_Q z^{-1}~
     \left( 1-\frac{1}{z}-\frac{\epsilon_Q}{1-z} \right)^{-2}~~,
\ee
where $N_Q$ is a normalization constant, and $Q$ denotes $b$ or $c$ quark. 
This functional form is not purely ad hoc., but motivated
by general theoretical arguments,
that the transition amplitude for a fast moving heavy quark $Q$
to fragment into a heavy meson $(Qq)$ is proportional to 
the inverse of the energy transfer $\Delta E^{-1}$.

This function (3) has a peak at $z \sim 1- \sqrt{\epsilon_Q}$.
Since the parameter $\epsilon_Q$ is related to the  ratio
of the effective light quark mass to the heavy quark mass 
$\sim m_q^2/m_Q^2$,
the peak approaches to 1 in the limit of $m_Q \ra \infty$, and
the function shows the similar behavior to $\delta(1-z)$.
Hence we find that this functional form is supported by the HQET
in this indirect manner.

In the Drell-Yan process, the rest degrees of freedom of initial
nucleons which do not take part in the scattering
make incoherent final states, see in Fig. 1 (a).
In the exclusive semileptonic decay of a heavy meson into a final state
heavy meson, however, two sets of left-over light-degrees of freedom are
summed to have the connection, i.e. see Fig. 2 (b).
To connect them we need a relation between $x$ and $y$
from the decay kinematics.
The momentum transfer of the decay between mesons is defined as
\be
q \equiv p_B - p_D~~.
\ee
On the other hand, the momentum transfer of the partonic subprocess
is given by
\be
q^{\mbox{{\tiny (parton)}}} &=& p_b - p_c \nonumber \\
                            &=& x p_B - y p_D~~.
\ee
Note that the second line of the Eq. (5) holds only  
in the infinite momentum frame while Eq. (4) holds at any frames.
In fact, the heavy meson's momentum would be $p_H=p_Q+k+{\cal O}
(1/m_Q)$, where $H=B,D$ or $D^*$, and  $Q=b$ or $c$. 
And $k$ denotes the momentum of the light-degrees of freedom,
and has the size of the effective mass 
of a common light degree of freedom, $\bar \Lambda$.
Therefore we have $q = q^{\mbox{{\tiny (parton)}}}$ up to
the common part of the $1/m_Q$ corrections 
in the infinite momentum frame, 
and with these kinematic relations, (4) and (5), 
we derive the following relation 
\be
y(x,q^2) = \frac{1}{m_D} \sqrt{ x(x-1) m_B^2 + (1-x) q^2 + x m_D^2}~~.
\ee
Note that if $x=1$, $y=1$, this equation describes the decay of
the free quark with mass $m_B$ to the final quark with mass $m_D$.
Substituting $y$ of Eq. (1) for $y(x, q^2)$ of Eq. (6), the double integral 
of Eq. (1) is reduced to the single integral over $x$.
Using this relation, we can sum the intermediate states 
in Fig. 1(b). 

We note here that the kinematic relation (6) are
valid approximations for the heavy-to-heavy resonance decays,
with the common light-degrees of freedom of the size ${\cal O}(1/m_Q)$.
As explained before, in the limit where $f_Q(x)= \delta(1-x)$ by increasing
$m_Q$ to infinity, we can reproduce the HQET leading term.
By comparison, the inclusive parton model approach is more reliable  for
the heavy-to-light non-resonant decays to final states of many particles.

\section{ Form Factors for semileptonic $B \ra D (D^*)$ Decays}

\subsection{$B \ra D e \nu$}

{}From Lorentz invariance we write the matrix element of the decay 
$\bar{B} \ra D e \bar{\nu}$ in the form
\be
<D|J_{\mu}|B> = f_+(q^2) (p_B+p_D)_{\mu} + f_-(q^2) (p_B-p_D)_{\mu}~~, 
\ee
and in terms of the HQET 
\be
<D(v')|J_{\mu}|B(v)> = \sqrt{m_B m_D}~(\xi_+(v \cdot v') (v+v')_{\mu} 
				     + \xi_-(v \cdot v') (v-v')_{\mu})~~.
\ee
Due to the conservation of leptonic currents, the form factors
multiplicated by $q_{\mu}$ do not contribute.
Then the hadronic tensor is given by
\be
H_{\mu \nu} &=& <D|J_{\mu}|B><B|J^{\dagger}_{\nu}|D>
           \nonumber \\
            &=& 2~|f_+(q^2)|^2 ({p_B}_{\mu} {p_D}_{\nu} 
                              + {p_B}_{\nu} {p_D}_{\mu})~~,
\ee
and can be expressed by the Isgur-Wise function,
\be
H_{\mu \nu} = R^{-1} |\xi(v \cdot v')|^2  
             ({p_B}_{\mu} {p_D}_{\nu} + {p_B}_{\nu} {p_D}_{\mu})
                       \left( 1+{\cal O}(\frac{1}{m_Q}) \right)~~,
\ee
where 
\be
R=\frac{2\sqrt{m_B m_D}}{m_B+m_D}~~. \nonumber
\ee

The partonic subprocess decay width for $B \ra D e \nu$ decay
is given by
\be
E_b \cdot d\Gamma (b \ra ce\nu) &=& 
             \frac{1}{2} (2\pi)^4 
             \delta^4(p_b-p_c-q^{\mbox{{\tiny (parton)}}} ) \cdot
            2 {G_F}^2 |V_{cb}|^2 H_{\mu \nu}^{\mbox{{\tiny (parton)}}} 
                                                L^{\mu \nu} \nonumber \\
       && ~~~~~~~~~~~\times
              \frac{d^3p_c}{(2\pi)^3 2E_c}
               \frac{d^3p_e}{(2\pi)^3 2E_e}
                \frac{d^3p_{\nu}}{(2\pi)^3 2E_{\nu}}~~,
\ee
where
\be
&&L^{\mu \nu} = 2~(p_e^{\mu} p_{\nu}^{\nu} + p_e^{\nu} p_{\nu}^{\mu} 
                -g^{\mu \nu} p_e \cdot p_{\nu}
                +i \epsilon^{\mu \nu \alpha \beta} 
                               {p_e}_{\alpha} {p_{\nu}}_{\beta})~~,
\nonumber \\
&&H_{\mu \nu}^{\mbox{{\tiny (parton)}}} 
           = N~({p_b}_{\mu} {p_c}_{\nu} + {p_b}_{\nu} {p_c}_{\mu})~~.
\ee
Here $H_{\mu \nu}^{\mbox{{\tiny (parton)}}}$ denotes the partonic 
subprocess'
hadronic tensor contributing $B \ra D$ decay. 
Now we need to discuss about it.
In the inclusive decays $B \ra X_c e \nu$, 
the hadronic tensor is given by 
\be
W_{\mu \nu}=2~({p_b}_{\mu} {p_c}_{\nu} + {p_b}_{\nu} {p_c}_{\mu} 
              -g_{\mu \nu} p_b \cdot p_c 
          -i \epsilon_{\mu \nu \alpha \beta} p_b^{\alpha} p_c^{\beta})~~. 
\nonumber
\ee
As can be easily seen, this hadronic tensor contains 
all the possible spin configurations of $b \ra c$ transition 
from spin $0$ state.
Here we are interested in only $B \ra D$ process, 
where the spin does not change during the process. 
Therefore, we have to choose only the spin-inert part 
which contributes to $B \ra D$ process from the inclusive hadronic tensor, 
~$W_{\mu \nu}$.
By comparing with Eq. (10), we find that the spin-inert part
has the form of $({p_b}_{\mu} {p_c}_{\nu} + {p_b}_{\nu} {p_c}_{\mu})$,
as shown in (12).
Besides, we do not know how much spin-inert part 
really contributes to $B \ra D$ semileptonic decay, because other decays
such as $B \ra D^*$, $B \ra D^{**}$ also contain spin-inert parts. 
Therefore, the parameter $N$ is introduced to estimate the size of 
spin-inert part which contributes to $B \ra D$ process. 
The constant $N$ will be later determined 
by the zero recoil limit of the Isgur-Wise function.

Using the relation (2), we can write the hadronic tensor in the parton 
level as follows 
\be
H_{\mu \nu}^{\mbox{{\tiny (parton)}}} 
           &=& N~xy~({p_B}_{\mu} {p_D}_{\nu} + {p_B}_{\nu} {p_D}_{\mu})~~.
\ee
The momentum conservation of the partonic  subprocess corresponds
to the momentum conservation in the hadronic level in our model, 
as explained before.
So we can substitute the Dirac delta function 
$ \delta^4(p_b-p_c-q^{\mbox{{\tiny (parton)}}} )$
for $ \delta^4(p_B-p_D-q)$ in Eq. (11) 
with no loss of generality.
Therefore, we write the decay width of $\bar{B} \ra De\bar{\nu}$,
\be
E_B \cdot d\Gamma (B \ra De\nu) 
          &=& \int~dx~f_B(x)~f_D(y(x,q^2)) E_b~d\Gamma(b \ra ce\nu)
                \nonumber \\
          &=& \int dx~f_B(x) f_D(y(x,q^2))~ 
                (2\pi)^4 \delta^4(p_B-p_D-q)
                \nonumber \\
          &&~~~~~~~~\times {G_F}^2 |V_{cb}|^2 
           ~N~xy(x,q^2)~({p_B}_{\mu} {p_D}_{\nu} + {p_B}_{\nu} {p_D}_{\mu}) 
                       L^{\mu \nu}
                \nonumber \\
          &&~~~~~~~~~~~~\times~y^2(x,q^2)~\frac{d^3p_D}{(2\pi)^3 2E_D}
                          \frac{d^3p_e}{(2\pi)^3 2E_e}
                          \frac{d^3p_{\nu}}{(2\pi)^3 2E_{\nu}}~~.
\ee
Hence we find the hadronic tensor
\be
H_{\mu \nu}(B \ra D e \nu)
         &=& N \int dx~f_B(x) f_D(y(x,q^2)) ~xy^3(x,q^2)~
               ({p_B}_{\mu} {p_D}_{\nu} + {p_B}_{\nu} {p_D}_{\mu})
               \nonumber \\
         &\equiv& N~{\cal F}(q^2) 
               ({p_B}_{\mu} {p_D}_{\nu} + {p_B}_{\nu} {p_D}_{\mu})~~,
\ee
where we factorize the function ${\cal F}(q^2)$ defined as
\be
{\cal F}(q^2) \equiv  \int dx~f_B(x) f_D(y(x,q^2)) ~xy^3(x,q^2)~~.
\ee
For given $q^2$ in our parton picture, the function ${\cal F}(q^2)$ 
measures the weighted transition amplitude, which is explicitly 
related to the overlap integral
of distribution functions of initial and final state hadrons.

Comparing (15) with the Eq. (10), the Isgur-Wise function is calculated 
within the parton model approach
\be
|\xi(v \cdot v')|^2 \left( 1 + {\cal O}(\frac{1}{m_Q}) \right)
                 = 4 N \cdot R \cdot {\cal F}(v \cdot v')~~.
\ee
As explained before, in our model the non-perturbative QCD effects are 
included through the distribution functions, so the function 
${\cal F}(q^2)$ contains all orders of $1/m_Q$ corrections.
Note that, however, our model ignores the effects of the transeverse 
quark motion and off--mass shell, the full $1/m_Q$ effects
are not being considered.
{}From the value of the zero recoil limit of the Isgur-Wise function 
with $1/m_Q$ corrections \cite{update}, we can determine the numerical 
value of $N$. In Table 1, we show the numerical values 
of $N$ with varying the parameters of the fragmentation functions, 
($\epsilon_b$, $\epsilon_c$). 
Unfortunately we cannot obtain the analytic form of
${\cal F}(q^2)$ due to the nontriviality of the integrals of $f(x)$.
However, we can numerically obtain the behaviors of the Isgur-Wise function
with varying $q^2$ from the Eq. (17), 
and calculate its slope parameter.
The normalization constant $N$ and the slope parameters $\rho^2$ 
with the input values of ($\epsilon_b$, $\epsilon_c$)
are also shown in the Table 1. 
We varied $\epsilon_b$ from 0.004 to 0.006, and
$\epsilon_c$ from 0.04 to 0.1. As explained earlier, 
the parameter $\epsilon_Q$ is related to the  ratio
of the effective light quark mass to the heavy quark mass 
$\sim m_q^2/m_Q^2$. And therefore, our input values of
($\epsilon_b$, $\epsilon_c$) correspond to the range
$m_c^2/m_b^2 = 0.04 \sim 0.15$, which is consistent with the 
prediction of the HQET \cite{neubert}. 
For more details on the comparison with the HQET, see Section IV.

The $q^2$ spectrum is given by
\be
\frac{d\Gamma(\bar{B} \ra D e \bar{\nu})}{dq^2} 
  = \frac{{G_F}^2 |V_{cb}|^2}{96 \pi^3 m_B^3} 
{\cal F}(q^2) N \left( (m_B^2-m_D^2+q^2)^2-4 m_B^2 q^2 \right)^{3/2}~~.
\ee
And in Fig. 2(a), our prediction is shown  
with the parameters ($\epsilon_b =$0.004,  $\epsilon_c =$0.04), 
compared with Wirbel {\it et al}.'s 
model prediction \cite{wsb}, which shows quite a good agreement.

\subsection{$B \ra D^* e \nu$}

We write the matrix element of the decay $\bar{B} \ra D^* e \bar{\nu}$
in familiar form
\be
<D^*|V_{\mu}+A_{\mu}|B>&&= 
           \frac{2i}{m_B+m_{D^*}} \epsilon_{\mu \nu \alpha \beta} 
           {\epsilon^*}^{\nu} p_{D^*}^{\alpha} p_B^{\beta} V(q^2)
        \nonumber \\
        + (m_B&&+m_{D^*}) {\epsilon^*}_{\mu} A_1(q^2)
    - \frac{\epsilon^* \cdot q}{m_B+m_{D^*}} (p_B+p_{D^*})_{\mu} A_2(q^2)
        \nonumber \\
        & & -2 m_{D^*} \frac{\epsilon^* \cdot q}{q^2} q_{\mu} A_3(q^2) 
         +2 m_{D^*} \frac{\epsilon^* \cdot q}{q^2} q_{\mu} A_0(q^2)~~,
\ee
where 
\be
A_3(q^2) = \frac{m_B+m_{D^*}}{2 m_{D^*}} A_1(q^2)
         - \frac{m_B-m_{D^*}}{2 m_{D^*}} A_2(q^2)~~, \nonumber
\ee
and in terms of the HQET
\be
<D^*(v')|V_{\mu}+A_{\mu}|B(v)> &&= \sqrt{m_Bm_{D^*}} 
           \nonumber \\
         \times ( i \xi_V(&&v \cdot v') \epsilon_{\mu \nu \alpha \beta} 
           {\epsilon^*}^{\nu} v'^{\alpha} v^{\beta} v(q^2)
           +\xi_{A_1}(v \cdot v')(v \cdot v'+1) {\epsilon^*}_{\mu}
           \nonumber \\
         &&-\xi_{A_2}(v \cdot v')\epsilon^* \cdot v v_{\mu}
           -\xi_{A_3}(v \cdot v')\epsilon^* \cdot v v'_{\mu})~~.
\ee
In fact, $A_0(q^2)$ and $A_3(q^2)$ do not
contribute here because of leptonic currents conservation.
With the heavy quark expansion, the hadronic form factors are related
to the Isgur-Wise function $\xi(v \cdot v')$, such as
$\xi_i(v \cdot v') = \xi(v \cdot v')\big( \alpha_i+{\cal O}(1/m_Q) \big)$,
where $\alpha_{A_2}=0$ and $\alpha_i=1$ otherwise.

The hadronic tensor of $\bar{B} \ra D^* e \bar{\nu}$ decay is obtained
using the HQET
\be
H_{\mu \nu} &=& \sum_{\mbox{{\tiny spin}}} 
             <D^*| V_{\mu}+A_{\mu} |B> <D^*| V_{\nu}+A_{\nu} |B>^*
             \\
            &=& {R^*}^{-1} |\xi(v \cdot v')|^2  
             \left[
	     \left(1-\frac{2q^2}{(m_B+m_{D^*})^2}\right)
	     ({p_B}_{\mu} {p_{D^*}}_{\nu} + {p_B}_{\nu} {p_{D^*}}_{\mu})
             (1+{\cal O}(\frac{1}{M_Q})) \right.
             \nonumber \\
      && \left.-2 \left( 1-\frac{q^2}{(m_B+m_{D^*})^2} \right)
             \left(g_{\mu \nu} p_B \cdot p_{D^*}
             (1+{\cal O}(\frac{1}{M_Q}))
          -i \epsilon_{\mu \nu \alpha \beta} p^{\alpha}_B p^{\beta}_{D^*}
             (1+{\cal O}(\frac{1}{M_Q})) \right)
             \right]~~,
             \nonumber 
\ee
where $ R^*=2\sqrt{m_B m_{D^*}}/(m_B+m_{D^*})$. \nonumber
Investigating the above relation, we can find the form of the 
hadronic tensor for the partonic subprocess which contributes to the decay 
$\bar{B} \ra D^* e \bar{\nu}$.
The resulting hadronic tensor is written down in the form,
\be
H_{\mu \nu}^{(\mbox{{\tiny parton}})} &=& 
                \left( 1-\frac{2q^2}{(m_B+m_{D^*})^2} \right) N_1~
	   ({p_b}_{\mu} {p_c}_{\nu} + {p_b}_{\nu} {p_c}_{\mu})
             \nonumber \\
          &&~~~~-2~\left( 1-\frac{q^2}{(m_B+m_{D^*})^2} \right)
           (N_2~g_{\mu \nu} p_b \cdot p_c 
     -i N_3~\epsilon_{\mu \nu \alpha \beta} p^{\alpha}_b p^{\beta}_c)~~.
\ee
The parameters $N_i$'s give the relative size of form factors and 
overall normalization. In general they are not constants and 
have the $q^2$ dependences.
In the heavy quark limit, $N_i$'s become
constants and the values are equal to that of the normalization
constant $N$ in $B \ra D e \nu$ process.

In order to investigate the procedure more conveniently, 
we define the ratios of form factors as follows:
\be
R_1 &\equiv& \left( 1-\frac{q^2}{(m_B+m_{D^*})^2} \right) 
             \frac{V(q^2)}{A_1(q^2)}~~,
             \nonumber \\
R_2 &\equiv& \left( 1-\frac{q^2}{(m_B+m_{D^*})^2} \right) 
             \frac{A_2(q^2)}{A_1(q^2)}~~,
\ee
where $V(q^2)$, $A_1(q^2)$ and $A_2(q^2)$ denote vector and axial vector
form factors respectively. 
Then we can write the relations among form factors 
and the Isgur-Wise function as
\be
A_1(q^2) &=& \left( 1-\frac{q^2}{(m_B+m_{D^*})^2} \right) 
             {R^*}^{-1} \xi(q^2)~~,
             \nonumber \\
A_2(q^2) &=& R_2 {R^*}^{-1} \xi(q^2)~~,
             \nonumber \\
V(q^2) &=& R_1 {R^*}^{-1} \xi(q^2)~~.
\ee
Note that the Isgur-Wise function $\xi(v \cdot v')$
in these expressions contains full $1/m_Q$ corrections and 
it corresponds to ${\hat \xi}(v \cdot v')$ in the Ref. \cite{update},
which is normalized at zero recoil up to the second order corrections
${\hat \xi}(1) = 1+\delta_{1/m_Q}$.
And $R_1$ and $R_2$ become to be unity in the heavy quark limit, and
Neubert's estimates of the $1/m_Q$ corrections  
\cite{neubert} give
\be
R_1 \approx 1.3~~,~~~~~~~~R_2 \approx 0.8~~,
\ee
which are model dependent.
Recently CLEO \cite{cleo2} obtained the values of the parameters 
$R_1$, $R_2$ 
by fitting them with the slope parameter of the Isgur-Wise function
$\rho^2$.
Since $R_1$ and $R_2$ have very weak $q^2$ dependence,  
in the CLEO analysis they approximately obtained the  $q^2$ 
independent values,
\be
R_1 &=& 1.30 \pm 0.36 \pm 0.16~~, \nonumber \\
R_2 &=& 0.64 \pm 0.26 \pm 0.12~~.
\ee
Hereafter we also take them to be constants for simplicity.

In our model the parameters $N_1$, $N_2$ and $N_3$ are represented in 
terms of $R_1$ and $R_2$ as follows,
\be
&&N_1 = \frac{N/2}{1-2q^2/(m_B+m_{D^*})^2} 
        \left[
        -\frac{q^2}{(m_B+m_{D^*})^2} \cdot 2R_1^2 \right.
          \nonumber \\
&& \left.~~ +\left( 1-\frac{q^2}{(m_B+m_{D^*})^2}\right)
          \left(
          \frac{(1+R_2)^2}{2} + \frac{m_B^2+q^2}{2m_{D^*}^2} (1-R_2)^2
          +\frac{2m_B m_{D^*}-q^2}{2m^2_{D^*}} (1-R_2^2)
          \right)
        \right]~~,
        \nonumber \\
&&N_2 = \frac{N}{2} \left[
        (1+R_1^2) + \frac{2m_B m_{D^*}}{m_B^2+m_{D^*}^2-q^2} (1-R_1^2)
        \right]~~,
        \nonumber \\
&&N_3 = N R_1~~.
\ee
In the heavy quark limit, we know that $R_1=R_2=1$.
Using the expression $ R_i = 1 + {\cal O}(1/m_Q) $
we can separate the leading contributions and $1/m_Q$
corrections in $N_i$'s:
\be
N_1 &=& \frac{N}{4} (1+R_2)^2 + {\cal O}(1-R_2)~~,
        \nonumber \\
N_2 &=& \frac{N}{2} (1+R_1^2) + {\cal O}(1-R_1)~~,
        \nonumber \\
N_3 &=& N R_1~~.
\ee
When $R_1,~R_2 \ra 1$, we explicitly see that $N_1=N_2=N_3 \ra N$.

Now substituting $\xi(v \cdot v')$ for ${\cal F}(q^2)$ with the Eq. (17),
the hadronic tensor for the decay $\bar{B} \ra D^* e \bar{\nu}$
is given by
\be
H_{\mu \nu}(B \ra D^*) &=& {\cal F}(q^2) 
         \left[
         \big(1-\frac{2q^2}{(m_B+m_{D^*})^2}\big) N_1
         ({p_B}_{\mu} {p_{D^*}}_{\nu} + {p_B}_{\nu} {p_{D^*}}_{\mu})
         \right.
             \nonumber \\
         &-& \left.2 \big(1-\frac{q^2}{(m_B+m_{D^*})^2}\big)
           (N_2 g_{\mu \nu} p_B \cdot p_{D^*} 
           -i N_3 \epsilon_{\mu \nu \alpha \beta} 
                     p^{\alpha}_B p^{\beta}_{D^*})
         \right]~~.
\ee
The $q^2$ dependences of form factors are mainly determined
by the function ${\cal F}(q^2)$, instead of the commonly used 
pole-dominance ans\"atze.

When we calculate the hadronic tensor within the HQET framework, 
we have generally some parameters parametrizing non-perturbative effects,
which are obtained in model dependent ways. 
The slope parameter is such a characteristic parameter of
the Isgur-Wise function, which
represents the common behaviors of form factors. 
We calculated it, and find that the value of the slope parameter 
is related to the parameters $\epsilon_b$ and $\epsilon_c$ 
in our approach. The HQET also contains the parameter 
$\lambda_1 \sim -\langle k_Q^2 \rangle$
which is related to the kinetic energy of the heavy quark 
inside the heavy meson \cite{cskim}, and spin-symmetry breaking term
$\lambda_2 = \frac{1}{4}(m_V^2-m_P^2)$
corresponding to the mass splitting of pseudoscalar mesons and vector mesons. 
In our approach, we have two parameters $R_1$ and $R_2$ correspondingly, 
which are experimentally measurable.
Also the values of $\epsilon_b$ and $\epsilon_c$ can be independently 
determined from the various methods experimentally and theoretically.
We use the values for $R_1$ and $R_2$ from the CLEO
fit results of Ref. \cite{cleo2}, and for $\epsilon_b$ and $\epsilon_c$
from Ref. \cite{peterson,zphy}.

Finally, we obtain the decay spectrum
\be
\frac{d\Gamma(\bar{B} \ra D^* e \bar{\nu})}{dq^2} 
  &=& \frac{G_F^2 |V_{cb}|^2}{192 \pi^3 m_B^5} {\cal F}(q^2) 
      \left( (m_B^2-m_{D^*}^2+q^2)^2-4 m_B^2 q^2 \right)^{1/2}
      \nonumber \\
  & & ~~~
      \times \left[
        m_B^2 W_1(q^2)~((m_B^2-m_{D^*}^2+q^2)^2 -m_B^2 q^2) \right.
      \nonumber \\
  & & ~~~~~~~
      \left. + \frac{3}{2} m_B^2 W_2(q^2) ~(m_B^2-m_{D^*}^2+q^2) 
                 + 3 m_B^2 W_3(q^2) 
      \right]~~,
\ee
where
\be
W_1(q^2) &=& - N_1 \left( 1-\frac{2q^2}{(m_B+m_{D^*})^2} \right)~~,
        \nonumber \\
W_2(q^2) &=&  N_1 (m_B^2-m_{D^*}^2+q^2) 
               \left(1-\frac{2q^2}{(m_B+m_{D^*})^2}\right)
       - 2 N_3 q^2 \left( 1-\frac{q^2}{(m_B+m_{D^*})^2} \right)~~,
        \nonumber \\
W_3(q^2) &=& - N_1 m_B^2 q^2 \left( 1-\frac{2q^2}{(m_B+m_{D^*})^2} \right)
        \nonumber \\
    & &~~~~+ N_3 q^2 (m_B^2-m_{D^*}^2+q^2) 
         \left( 1-\frac{q^2}{(m_B+m_{D^*})^2} \right)
        \nonumber \\
    & &~~~~~~~~+ N_2 q^2 (m_B^2+m_{D^*}^2-q^2) 
         \left( 1-\frac{2q^2}{(m_B+m_{D^*})^2} \right)~~,
\ee
and ${\cal F}(q^2)$ is defined in (16).
The result is plotted in Fig. 2 (b), also
compared  with the CLEO data \cite{cleo2}.
The thick solid line is our model prediction 
with the parameters ($\epsilon_b =$0.004,  $\epsilon_c =$0.04) and
the dashed line the Wirbel {\it et al.} model prediction \cite{wsb}. 

\section{Discussions and Conclusions}

All form factors show the common behavior for
varying $q^2$, which is described by the Isgur-Wise 
function of the HQET, which represents the common non-perturbative
dynamics of weak decays of heavy mesons.
Ever since Fakirov and Stech \cite{fakirov}, the
nearest pole dominance has been usually adopted as the dependence of
common behaviors on $q^2$. 
In our approach, their $q^2$ dependences are  extracted
from the kinematic relations of $b$- and $c$-quark. 
When $b$-quark decays to $c$-quark, the momentum transfer 
to leptonic sector is equal to
the difference between $b$-quark momentum and $c$-quark momentum
in the parton picture.
The $b$- and $c$-quark momenta within the $B$ and $D$ mesons
have some specific distributions.
For given momentum transfer $q^2$, there exist possible
configurations of $b$- and $c$-quark momentum pairs $(p_b,~p_c)$, 
and each pair is appropriately weighted with the momentum distributions
of the quarks.
Our ${\cal F}(q^2)$ function in (16) measures the weighted transition 
amplitude for given $q^2$ in the parton picture; it is explicitly 
given by the overlap integral
of distribution functions of initial and final state hadrons.
This is common to all form factors, as explained in Section II.

As mentioned earlier, the non-perturbative strong interaction
effects are also considered through the distribution functions 
in our model, so $\xi(v \cdot v')$ obtained from Eq. (17)
corresponds to the hadronic form factor ${\hat \xi}(v \cdot v')$
defined in the Ref. \cite{update}, including $1/m_b$ corrections
rather than the lowest order Isgur-Wise function.
And the slope parameter of our results in Table 1 also corresponds to
${\hat \rho}^2$ related to ${\hat \xi}(v \cdot v')$.
We obtain the values of the slope parameter $\rho$ within the 
parton model framework, as in Table 1,
\be
 \rho^2 = 0.552 - 0.858~~,
 \nonumber
\ee
which are compatible with the Neubert's prediction \cite{update},
\be
{\hat \rho}^2 \simeq \rho^2 \pm 0.2 = 0.7 \pm 0.2~~.
 \nonumber
\ee
Our result is  rather smaller than the predictions of other models,
\be
 \rho^2 &=& 1.29 \pm 0.28 ~~~~~~\mbox{\cite{rosner}}~~,
 \nonumber \\
 \rho^2 &=& 0.99 \pm 0.04 ~~~~~~\mbox{\cite{mannel}}~~,
 \nonumber 
\ee
but it is consistent with the average value measured 
by experiments \cite{exps},
\be
 {\hat \rho}^2 = 0.87 \pm 0.12~~.
 \nonumber
\ee

In calculating the numerical values, we still have two free
parameters $\epsilon_b$ and $\epsilon_c$ of Eq. (6),
{\it i.e.} of the heavy quark fragmentation functions. 
Their values can be determined 
independently from the various experimental and theoretical 
methods\footnote{Theoretically, it is interesting to calculate
$f_Q(z)=\delta(1-z)+{\cal O}(1/m_Q^2)$ within the HQET, which can reproduce
the phenomenological predictions.}.
For the parton model to be consistent with the HQET,
we require that with the fixed value of the parameter $\epsilon_Q$,
all the appropriate results of the parton model approach agree
with those of the HQET. 
In other words, the value of parameter $\epsilon_Q$ should be determined
to give all the phenomenological results to coincide
with those of the HQET.
In this point of view, we have previously studied the parton model
approach
for inclusive semileptonic decays of $B$ meson in the Ref. \cite{ky},
and showed that the value $\epsilon_b \approx 0.004$
gives consistent results with those of the HQET.
In this paper we use the value of $\epsilon_b$ as 0.004
or 0.006. The latter value is given by the experiments for the
determination of the Peterson fragmentation function \cite{zphy}.
We find that our prediction of the slope parameter $\rho^2$
with the parameter $\epsilon_b = 0.004$ and $\epsilon_c = 0.04$ 
gives the best agreed value $\sim 0.7$ 
with that of the HQET,
${\hat \rho}^2 = 0.7 \pm 0.2$. In this context, 
we conclude that our model with the parameter $\epsilon_b = 0.004$
gives consistent predictions with the HQET.

To investigate the possible model dependences of our approach
on the input fragmentation functions, 
we now study the cases with other fragmentation functions.
Following the Artru--Mennessier string model \cite{artru} for heavy 
flavors, we have the fragmentation function such as 
\be
f_Q(z) = N_Q \frac{(1-z)^a}{z^{1+bm_Q^2}} 
            \exp \left( -\frac{bm_Q^2}{z} \right)~~,
\ee
with parameters $b=0.8$ GeV$^{-2}$ and $a \approx 0.5$, and
the $c$-- and $b$--quark masses identified with the masses of the
lowest--lying vector meson states \cite{morris}.
Another fragmentation function is recently developed
with the help of the perturbative QCD (PQCD).
Inspired by the PQCD of the heavy quarkonium, the functional 
form for the heavy--light mesons \cite{pqcd} is derived as
\be
f_Q(z)&=&N_Q \frac{r_Q z(1-z)^2}{(1-(1-r)z)^6}
         \left( 6-18(1-2r_Q)z + (21-74r_Q+68r_Q^2)z^2 \right.
         \nonumber \\
      && ~~~~ \left. -2(1-r_Q)(6-19r_Q+18r_Q^2)z^3
                +3(1-r_Q)^2(1-2r_Q+2r_Q^2)z^4 \right)~~,
\ee
with the free parameter $r_Q$ being the mass relation
$m_Q/(m_Q+m_q)$, and with the light quark mass $m_q$.
We call this function the PQCD inspired fragmentation function.
With these two fragmentation functions, we obtain the numerical values of
$\rho^2 = 1.124$ from the string fragmentation function, 
and $\rho^2 = 0.605$ from the PQCD inspired fragmentation function.
The results are shown in Table 2 compared with those with
Peterson's function.
As can be seen, the string fragmentation function gives a somewhat 
larger value of $\rho$, but the `recently developed' PQCD inspired function 
gives a similar value to that of Peterson fragmentation function.

Phenomenologically, our model prediction on $q^2$ 
spectrum in the $B \ra D^* e \nu$ decay shows a good
agreement with the result of the CLEO \cite{cleo2}, as shown in Fig. 2(b).
If we let $f_Q(z)=\delta(1-z)$ and $R_1=R_2 =1$ in our model, 
we can reproduce the lowest order results of the HQET, and
obtain the similar plot with those of other models in Fig. 2(b).
For the $B \ra D e \nu$ decay, our results agree with those
of other models, as in Fig. 2(a).

Finally we obtain the ratio of integrated total widths
$\Gamma (B \ra D^*)/ \Gamma (B \ra D) \approx 2.66$,
which agrees with the experimental results \cite{pdg}.
It may be a phenomenological support of our model
because this quantity is independent of the CKM elements $|V_{cb}|$,
which has uncertainties in determining its value yet.
The perturbative QCD corrections can be factorized in the decay width
calculation \cite{update,pert},
which does not affect the ratio $\Gamma (B \ra D^*)/ \Gamma (B \ra D)$.

To summarize, we investigated the form factors for exclusive semileptonic 
decays of $B$-meson to $D,~D^*$, based on the parton picture 
and helped by the results of the HQET.  
We obtained the numerical results for the slope of the 
Isgur-Wise function, which is consistent
with the experimental results, and 
we extracted the dependences of hadronic form factors 
on $q^2$ by varying input heavy quark fragmentation function
without the nearest pole dominance ans\"atze. 


\vskip 1.0cm
\begin{center}
{\bf ACKNOWLEDGEMENTS}
\end{center}

We thank Pyungwon Ko and E. Paschos for their careful reading 
of manuscript and their valuable comments.
The work was supported 
in part by the Korean Science and Engineering  Foundation, 
Project No. 951-0207-008-2,
in part by Non-Directed-Research-Fund, Korea Research Foundation 1993, 
in part by CTP, Seoul National University, 
in part by Yonsei University Faculty Research Grant 1995, 
in part by the Basic Science Research Institute Program, Ministry of 
Education, 1997,  Project No. BSRI-97-2425, and in part by 
COE fellowship of Japanese Ministry of Education, Science and Culture.

%

\begin{table}
\begin{center}
\vspace{1cm}
\caption{
The values of the normalization constant $N$ and the slope parameter 
$\rho^2$ are shown with several choices for the input parameters 
$\epsilon_b$ and $\epsilon_c$.
}
\vspace{2cm}
\begin{tabular}{cccccccccccccr}
\hline
\hline
&&&&&&&&&&&&&\\
& & & & $\epsilon_b =$ & 0.004 & & & & $\epsilon_b =$ & 0.006 & & &  \\  
& & & $\epsilon_c =$ 0.04 & 0.06 & 0.08 &0.1 & & 
0.04 & 0.06 & 0.08 & 0.1 &\\
\hline 
&&&&&&&&&&&&&\\
& $N$ & & 0.938 & 1.068 & 1.190 & 1.306 & & 1.105 & 1.223 & 1.338 & 1.449 &
\\
&&&&&&&&&&&&&\\
& $\rho^2$ & & 0.705 & 0.646 & 0.609 & 0.582 & & 0.896 & 
0.844 & 0.810 & 0.785 &
\\
&&&&&&&&&&&&&\\
\hline
\hline
\end{tabular}
\vspace{2.5cm}
\caption{
Comparison of the slope parameter $\rho^2$ 
for the different fragmentation functions
}
\vspace{2cm}
\begin{tabular}{ccccccr}
\hline
\hline
&&&&&&\\
& Fragmentation Function (F. F.) & & & slope parameter $\rho^2$ & &  \\  
&&&&&&\\
\hline 
&&&&&&\\
& Peterson F. F. & & & 0.5521 -- 0.8582 & &  \\  
&&&&&&\\
& String F. F. & & & 1.1240 & &  \\  
&&&&&&\\
& PQCD inspired F. F. & & & 0.6053 & &  \\  
&&&&&&\\
\hline
\hline
\end{tabular}
\end{center}
\end{table}
%
%

\newpage
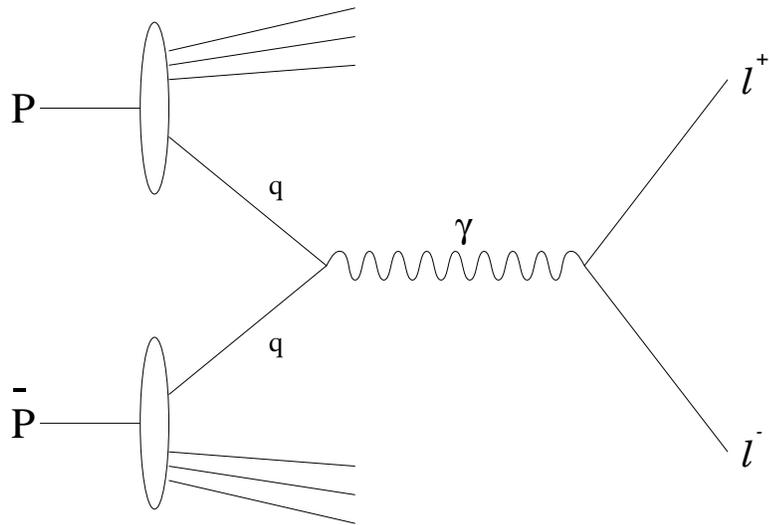
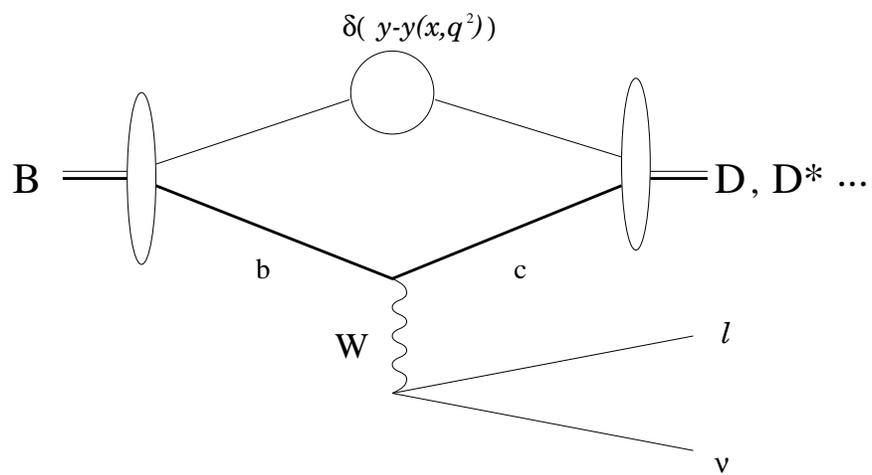
\begin{figure}
\caption{
(a) The diagram of Drell-Yan Process. (b) The schematic diagram of 
semileptonic exclusive decay of $B$-meson in the parton model.  }
\end{figure}

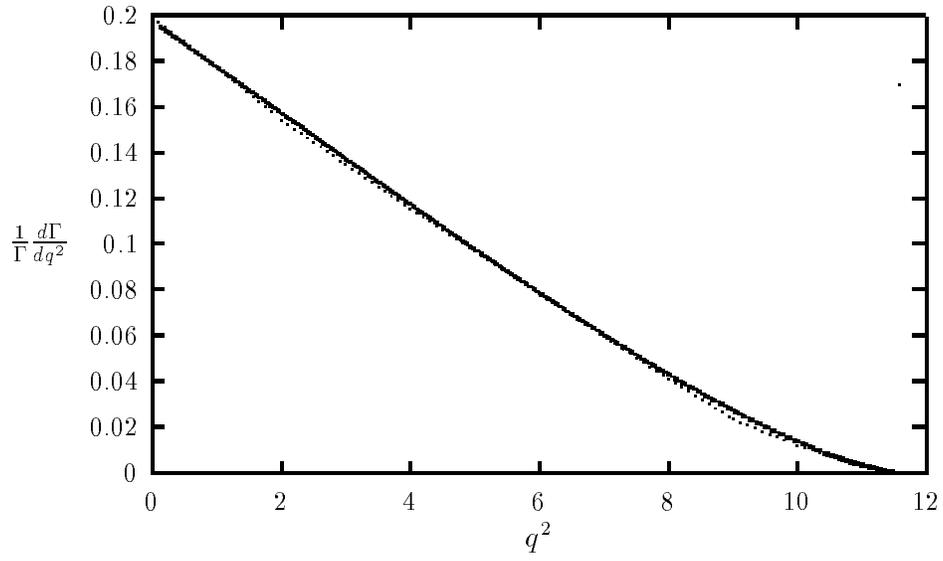
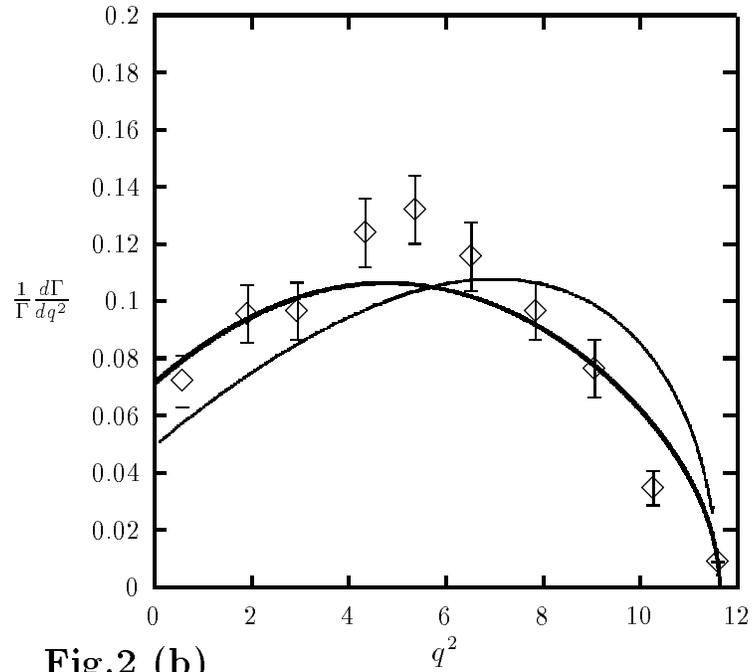
\begin{figure}
\caption{
(a) $q^2$ spectrum in the $B \ra D e \nu$ decays. 
The solid line is our model prediction and
the dotted line the Wirbel {\it et al.} model prediction \protect\cite{wsb}.
We used the values of parameters,
($\epsilon_b =$0.004,  $\epsilon_c =$0.04).
(b) $q^2$ spectrum in the $B \ra D^* e \nu$ decays. 
The solid line is our model prediction and
the dotted line the Wirbel {\it et al.} 
model prediction \protect\cite{wsb}.
We used the values of parameters,
($\epsilon_b =$0.004,  $\epsilon_c =$0.04).
The data are quoted from Ref. \protect\cite{cleo2}.
}
\end{figure}

\end{document}